\documentstyle[12pt]{article}

\def\Journal#1#2#3#4{{#1} {\bf #2}, #3 (#4)}
\def\PLB{{\em Phys. Lett.}B}

\def\PRD{{\em Phys. Rev.} D}

\newcommand{\beq}{\begin{equation}}
\newcommand{\eeq}{\end{equation}}

\def\sst{\scriptscriptstyle}

\def\be{\begin{equation}}
\def\ee{\end{equation}}
\def\bea{\begin{eqnarray}}
\def\eea{\end{eqnarray}}

\begin{document}
\begin{titlepage}
\begin{flushright}
\hfill{YUMS 98-002}\\
\hfill{SNUTP 98-002}\\
\end{flushright}
\vspace{1.5cm}
\begin{center}
{\large \bf USING $B_d^0,~B_s^0$ DECAYS AT HADRONIC $B$-FACTORIES 
TO DETERMINE CP ANGLES $\alpha$ AND $\beta$\footnote{
Talk given at the APCTP Workshop on Pacific Particle Physics Phenomenology, 
Oct 31 -- Nov 2, 1997. Seoul Korea
}}\\
\vspace{0.5cm}
{\bf C.S. Kim }\\
\vspace{0.5cm}
{\it Department of Physics, Yonsei University, Seoul 120-749, Korea }\\
\end{center}
%%%%%%%%%%%%%%%%%%%%%%%%%%%%%%%%
%
% You may repeat \author \address
% as often as necessary
%
%%%%%%%%%%%%%%%%%%%%%%%%%%%%%%%

%\maketitle

\begin{abstract}
At the first PPPP workshop, I gave a review talk on physics of
$B$ and CP-phases. In my talk, I explained DRG method and our extension to
extract CP angles $\alpha$ and $\gamma$ from measurements of the decay rates of 
$B_{d,s}^0 \to \pi\pi,~\pi K,~K K$.
There are several advantages to this extension:
discrete ambiguities are removed, fewer assumptions are necessary, and the
method works even if all strong phases vanish.
I also talked on several other topics.
\end{abstract}

\end{titlepage}

\newpage

\section{Introduction}
Here I review an extension of DGR method \cite{DGR} which avoids
most of the problems with the DGR method. In addition to the $B_d^0$ and
$B^+$ decays used by DGR, it requires the measurement of their
$SU(3)$-counterpart $B_s^0$ decays: $B_s^0 \to \pi^+ K^-$, $B_s^0(t) \to K^+ K^-$,
and $B_s^0 \to K^0 {\bar K^0}$. This overconstrains the system, which
eliminates most discrete ambiguities in the extraction of $\alpha$
and $\gamma$.

Although challenging experimentally, this method can probably be carried
out at hadron-collider experiments such as BTeV and LHC-B \cite{SStone}.
All branching ratios are ${\cal O}(10^{-5})$, and all final states involve only
charged pions and $K$'s -- there are no $\pi^0$'s. It will be necessary to
resolve $B_s^0$-${\overline{B_s^0}}$ oscillations, 
but this should not be be a problem at
hadron colliders: for example, BTeV expects to be able to measure values of
the mixing parameter $x_s \sim 50$. The principal obstacle to obtaining
precise values of $\alpha$ and $\gamma$ is the fact that 12 measurements
must be combined to extract these CP angles, as well as other quantities.
If the experimental errors are large, then all sensitivity to the CP angles
may be washed out. However, preliminary studies indicate that the errors
will in fact be relatively small. A BTeV simulation of the decay
$B_d^0(t) \to \pi^+\pi^-$ has been carried out, with the result that the CP
asymmetry in this decay can be measured with an error of 
$\pm 0.05$ \cite{SStone}. 
This error includes the background, which consists of the
decays $B_d^0\to\pi K$, $B_s^0 \to \pi K$ and $B_s^0 \to KK$. Since these are
precisely the decays involved in our extension of the DGR method, this
suggests that these decays can be experimentally separated from one
another, and their branching ratios measured with reasonable precision, say
$\sim 5\%$ or less. If this is so, then we can expect that $\alpha$ and
$\gamma$ can also be extracted fairly precisely. 

\section{ Using $B^0_s$ decays to determine the CP angles $\alpha$ and $\beta$ }
The DGR method involves the decays $B_d^0 \to\pi^+\pi^-$, $B_d^0 \to \pi^-
K^+$, and $B^+ \to \pi^+ K^0$. The amplitudes for these decays can be
written
\bea
\label{PPamps}
A_{\pi\pi} &\equiv & A\left(B^0\rightarrow \pi^+\pi^-\right) = 
 - \left( T + P \right), \nonumber \\
A_{\sst \pi K} &\equiv & A\left(B^0\rightarrow \pi^- K^+\right) = 
 - \left( T^\prime + P^\prime \right), \\
A_{\sst \pi K}^+ &\equiv & A\left(B^+\rightarrow\pi^+K^0\right) = 
 P^\prime ~.
\nonumber
\eea
Note that, since only tree and penguin terms are involved, EWP
contributions are negligible. 

The weak phase of $T$ is Arg$\left(V_{ud} V_{ub}^* \right) = \gamma$, and
similarly for $T'$: Arg$\left(V_{us} V_{ub}^* \right) = \gamma$. The $b\to
s$ penguin $P'$ is dominated by the internal $t$-quark, so its weak phase is
Arg$\left(V_{ts} V_{tb}^* \right) = \pi$. As for the $b\to d$ penguin $P$, 
if it also is dominated by the $t$-quark, its weak phase is Arg$\left(V_{td}
V_{tb}^* \right) = - \beta$. This is the assumption made by DGR, but we
can relax it.

If $SU(3)$ were unbroken, the amplitudes $T$ and $T'$ would be related simply
by the ratio of their CKM matrix elements: $|T'/T| = |V_{us}/V_{ud}|$.
However, if one includes first-order $SU(3)$ breaking, there
is an additional factor involving the ratio of $K$ and $\pi$ decay constants
if factorization is assumed:
$| {T^\prime}/{T} | = { |V_{us}| f_{\sst K}} /
 { | V_{ud} | f_\pi } \equiv r_u $.
On the other hand, since factorization is unlikely to hold for penguin
amplitudes, $P$ and $P'$ are not related in a simple way. However, DGR do
assume that the strong phase of the penguin diagram, $\delta_{\sst P}$, is
unaffected by $SU(3)$ breaking. 

With these assumptions, the amplitudes in Eqs.~(\ref{PPamps}) can be
written
\bea
A_{\pi\pi} & = & {\cal T} e^{i \delta_T} e^{i\gamma } +
 {\cal P} e^{i \delta_P} e^{- i\beta } ~, \nonumber \\
A_{\sst \pi K} & = & r_u {\cal T} e^{i \delta_T} e^{i\gamma } -
 {\cal P}^\prime e^{i \delta_P} ~, \\
A_{\sst \pi K}^+ & = & {\cal P}^\prime e^{i \delta_P} ~, \nonumber
\eea
where ${\cal T} \equiv \left| T \right|$, ${\cal P} \equiv \left| P
\right|$, and ${\cal P}^\prime \equiv \left| P^\prime \right|$.

There are thus six unknown quantities in the above 3 amplitudes: $\alpha
\equiv \pi - \beta - \gamma$, $\gamma$, ${\cal T}$, ${\cal P}$, ${\cal
P}^\prime$, and $\delta \equiv \delta_{\sst T} - \delta_{\sst P}$. 
The time-dependent, tagged $B_d^0$
and ${\overline{B_d^0}}$ decay rates to $\pi^+\pi^-$ are given by
\bea
\Gamma\left( B_d^0(t) \rightarrow \pi^+ \pi^- \right) &=& 
 e^{-\Gamma t} \left[ 
 \left| A_{\pi \pi } \right|^2 \cos^2 
 \left( \frac{\Delta m}{2} t 
 \right) +
 \left| \bar{A}_{\pi \pi } \right|^2 \sin^2 
 \left( \frac{\Delta m}{2} t 
 \right) 
 \right. \nonumber \\ 
 \ &\ & \hspace{1.5cm} + \left. 
 {\rm Im}\left(e^{2i\beta }A_{\pi \pi }
 \bar{A}_{\pi \pi }^* \right)
 \sin(\Delta mt)
 \right], \nonumber \\ 
\Gamma\left( {\overline{B_d^0}}(t) \rightarrow \pi^+ \pi^- \right) &=& 
 e^{-\Gamma t} \left[ 
 \left| A_{\pi \pi } \right|^2 \sin^2 
 \left( \frac{\Delta m}{2} t 
 \right) +
 \left| \bar{A}_{\pi \pi } \right|^2 \cos^2 
 \left( \frac{\Delta m}{2} t 
 \right) 
 \right. \nonumber \\
 \ &\ & \hspace{1.5cm} - \left. 
 {\rm Im}\left(e^{2i\beta }A_{\pi \pi }
 \bar{A}_{\pi \pi }^* \right)
 \sin(\Delta mt)
 \right]. 
\eea
{}From these measurements one can determine the three quantities $\left|
A_{\pi \pi } \right|^2$, $\left| \bar{A}_{\pi \pi } \right|^2$, and 
${\rm Im}\left(e^{2i\beta} A_{\pi\pi} \bar{A}_{\pi \pi }^*\right)$. 
The rates for the self-tagging decays $B_d^0 \to \pi^- K^+$ 
and ${\overline{B_d^0}} \to \pi^+ K^-$
are
\bea
|A_{\sst \pi K}|^2 & = & r_u^2 {\cal T}^2 + 
{{\cal P}^\prime}^2 - 2 r_u {\cal T}
{\cal P}^\prime \cos (\delta + \gamma), \nonumber \\
|{\bar A}_{\sst \pi K}|^2 & = & r_u^2 {\cal T}^2 + 
{{\cal P}^\prime}^2 - 2 r_u
{\cal T} {\cal P}^\prime \cos (\delta - \gamma).
\eea
Finally, the rates for $B^+ \to \pi^+ K^0$ and its CP-conjugate decay give
\beq
|A_{\sst \pi K}^+|^2=|{\bar A}_{\sst \pi K}^-|^2={{\cal P}^\prime}^2 ~.
\eeq

Thus, from the above measurements, one can obtain the following six
quantities:
\bea
A &\equiv & \frac{1}{2}
 \left( 
 \left| A_{\pi \pi }\right|^2 
 + \left| \bar{A}_{\pi \pi }\right|^2 
 \right) = 
 {\cal T}^2 + {\cal P}^2 
 - 2 {\cal T}{\cal P} \cos\delta \cos\alpha, 
\label{A1} \\
B &\equiv & \frac{1}{2}
 \left( 
 \left| A_{\pi \pi }\right|^2 
 - \left| \bar{A}_{\pi \pi }\right|^2 
 \right) = 
 - 2 {\cal T}{\cal P} \sin\delta \sin\alpha, 
\label{B1} \\
C &\equiv & {\rm Im}\left( e^{2 i \beta} A_{\pi \pi } 
 \bar{A}_{\pi \pi }^* 
 \right)
 = - {\cal T}^2 \sin2\alpha 
 + 2 {\cal T P} \cos\delta \sin\alpha, 
\label{C1} \\
D &\equiv & \frac{1}{2}
 \left( 
 \left| A_{\sst \pi K}\right|^2 
 + \left| \bar{A}_{\sst \pi K}\right|^2 
 \right) = 
 r_u^2 {\cal T}^2 + {{\cal P}^\prime}^2 
 - 2 r_u {\cal T}{\cal P}^\prime
 \cos\delta \cos\gamma, 
\label{D1} \\
E &\equiv & \frac{1}{2}
 \left( 
 \left| A_{\sst \pi K}\right|^2 
 - \left| \bar{A}_{\sst \pi K}\right|^2 
 \right) = 
 2 r_u {\cal T}{\cal P}^\prime
 \sin\delta \sin\gamma, 
\label{E1} \\ 
F &\equiv & \left| A_{\sst \pi K}^+\right|^2 = {{\cal P}^\prime}^2~.
\label{F1}
\eea
These give 6 equations in 6 unknowns, so that one can solve for 
$\alpha$, $\gamma$, ${\cal T}$, ${\cal P}$, ${\cal P}^\prime$, and $\delta$.
However, because the equations are nonlinear, there are discrete
ambiguities in extracting these quantities. In fact, a detailed study
shows that, depending on the actual values of the phases, there
can be up to 8 solutions.

If $\delta=0$, the quantities $B$ and $E$ vanish, so that one is left with 4
equations in 5 unknowns. In this case one must use additional assumptions
to extract information about the CP phases. Furthermore, even if $\delta
\ne 0$, if one relaxes any of the assumptions described above, the method
breaks down. For example, if one allows the strong phase of the $P'$ diagram
to be different from that of the $P$ diagram, as might be the case in the
presence of $SU(3)$ breaking, then one has 6 equations in 7 unknowns. And
if one relaxes the assumption that the $b\to d$ penguin is dominated by
the $t$-quark, then once again additional parameters are introduced, and
the method breaks down.

All of these potential problems can be dealt with by considering
additional $B_s^0$ decays. 
The problems with the DGR method can be resolved by adding amplitudes
which depend on the same 6 quantities, thus overconstraining the system. In
this case, if one adds a parameter or two, perhaps by relaxing certain
assumptions, the method will be less likely to break down. 

Within $SU(3)$ symmetry, the obvious decays to consider are the $SU(3)$
counterparts to the DGR decays, namely $B_s^0 \to \pi^+ K^-$, $B_s^0(t) \to K^+
K^-$, and $B_s^0 \to K^0 {\bar K^0}$. The amplitudes for these decays are
completely analogous to those in Eqs.~(\ref{PPamps}):
\bea
\label{BsPPamps}
B_{\sst \pi K} &\equiv & A\left(B_s^0\rightarrow \pi^+ K^-\right) = 
 - \left( \tilde{T} + \tilde{P} \right), \nonumber \\
B_{\sst KK} &\equiv & A\left(B_s^0\rightarrow K^+ K^-\right) = 
 - \left( {\tilde T}^\prime + \tilde{P}^\prime \right), \\ 
B_{\sst KK}^s &\equiv & A\left(B_s^0\rightarrow K^0{\overline{K^0}} 
\right) = \tilde{P}^\prime ~.
\nonumber
\eea
Here we have denoted the tree and penguin diagrams involving a spectator
$s$ quark by $\tilde{T}$ and $\tilde{P}$, respectively. As before, the
unprimed and primed quantities denote $\Delta S=0$ and $\Delta S=1$
processes, respectively. 

The weak phase of $\tilde{T}$ and $\tilde{T}^\prime$ is $\gamma$, and that
of $\tilde{P}^\prime$ is $\pi$. As for $\tilde{P}$, as a first step we make
the same assumptions as DGR, namely that it is dominated by the $t$-quark,
so that its weak phase is $-\beta$. Turning to $SU(3)$ breaking, we assume
factorization for the tree amplitudes, so that
$| {\tilde{T}^\prime} / {\tilde{T}} | = r_u $.
The magnitudes of the $\tilde{P}$ and $\tilde{P}^\prime$ amplitudes are
unrelated to one another. However, again as a first step, like DGR we
assume that they have the same strong phase, $\delta_{\sst\tilde{P}}$. 
The one new assumption that we make is that the relative strong phase
between the tree and penguin amplitudes is independent of the flavor of the
spectator quark. Thus we have $\delta_s = \delta$, where $\delta_s \equiv
\delta_{\sst\tilde{T}} - \delta_{\sst\tilde{P}}$ and $\delta \equiv
\delta_{\sst T} - \delta_{\sst P}$. (The most likely way for this to occur
is if $\delta_{\sst T} = \delta_{\sst\tilde{T}}$ and $\delta_{\sst P} =
\delta_{\sst\tilde{P}}$.) This assumption is motivated by the
spectator model.

Under these assumptions, the amplitudes in Eqs.~(\ref{BsPPamps}) can be
written
\bea
B_{\sst \pi K} & = & \tilde{{\cal T}} e^{i \delta_T} e^{i\gamma } +
 \tilde{{\cal P}} e^{i \delta_P} e^{- i\beta } ~, \nonumber \\
B_{\sst KK} & = & r_u \tilde{{\cal T}} e^{i \delta_T} e^{i\gamma } -
 \tilde{{\cal P}}^\prime e^{i \delta_P} ~, \\
B_{\sst KK}^s & = & \tilde{{\cal P}}^\prime e^{i \delta_P} ~, \nonumber
\eea
where $\tilde{\cal T} \equiv \left| \tilde{T} \right|$, $\tilde{\cal P}
\equiv \left| \tilde{P} \right|$, and $\tilde{\cal P}^\prime \equiv \left|
\tilde{P}^\prime \right|$.

The important point here is that three new parameters have been introduced
in the above amplitudes: $\tilde{{\cal T}}$, $\tilde{{\cal P}}$, and
$\tilde{{\cal P}}^\prime$. However, as in the DGR method, 6 quantities can
be extracted from measurements of the rates for these decays. Here, the
self-tagging decays are $B_s^0 \to \pi^+ K^-$ and
 ${\overline{B_s^0}} \to \pi^- K^+$,
whose rates are
\bea
|B_{\sst \pi K}|^2 & = & {\tilde{\cal T}}^2 + 
{\tilde{\cal P}}^2 - 2 \tilde{\cal T} \tilde{\cal P} 
\cos (\delta - \alpha), \nonumber \\
|{\bar B}_{\sst \pi K}|^2 & = & 
{\tilde{\cal T}}^2 + {\tilde{\cal P}}^2 - 2
\tilde{\cal T} \tilde{\cal P} \cos (\delta + \alpha).
\eea
The time-dependent, 
tagged $B_s^0$ and ${\overline{B_s^0}}$ decay rates to $K^+K^-$ are
given by 
\bea
\Gamma\left[ B_s^0(t) \rightarrow K^+ K^- \right] &=& 
 e^{-\Gamma t} \left[ 
 \left| B_{\sst KK} \right|^2 \cos^2 
 \left( \frac{\Delta m_s}{2} t 
 \right) +
 \left| \bar{B}_{\sst KK} \right|^2 \sin^2 
 \left( \frac{\Delta m_s}{2} t 
 \right) 
 \right. \nonumber \\
 \ &\ & \hspace{1.5cm} + \left. 
 {\rm Im}\left( B_{\sst KK}
 \bar{B}_{\sst KK}^* \right)
 \sin(\Delta m_s t)
 \right], \nonumber \\ 
\Gamma\left[ {\overline{B_s^0}}(t) \rightarrow K^+ K^- \right] &=& 
 e^{-\Gamma t} \left[ 
 \left| B_{\sst KK} \right|^2 \sin^2 
 \left( \frac{\Delta m_s}{2} t 
 \right) +
 \left| \bar{B}_{\sst KK} \right|^2 \cos^2 
 \left( \frac{\Delta m_s}{2} t 
 \right) 
 \right. \nonumber \\ 
 \ &\ & \hspace{1.5cm} - \left. 
 {\rm Im}\left( B_{\sst KK}
 \bar{B}_{\sst KK}^* \right)
 \sin(\Delta m_s t)
 \right], 
\eea
from which the quantities $\left| B_{\sst KK} \right|$, $\left|
\bar{B}_{\sst KK} \right|$, and ${\rm Im}\left( B_{\sst KK}\bar{B}_{\sst
KK}^* \right)$ can be extracted. Finally, we turn to $B_s^0(t) \to K^0
{\overline{K^0}}$. In principle there can be indirect CP violation in
these decays. However, within the SM, this CP violation is zero to a good
approximation, 
since both $B_s^0$-${\overline{B_s^0}}$ mixing and the $b\to s$ penguin
diagram, which dominates this decay, are real. Thus, measurements of the
rates for these decays yield
\beq
|B_{\sst KK}^s|^2 = |{\bar B}_{\sst KK}^s|^2 = 
{{\tilde{\cal P}}^\prime}{}^2 ~.
\eeq
Obviously, any violation of this equality will be clear evidence for new
physics.

Therefore the above measurements yield 6 new quantities:
\bea
\tilde{A} &\equiv & \frac{1}{2}
 \left( 
 \left| B_{\sst \pi K}\right|^2 
 + \left| \bar{B}_{\sst \pi K}\right|^2 
 \right) = 
 \tilde{\cal T}^2 + \tilde{\cal P}^2 
 - 2 \tilde{\cal T}
 \tilde{\cal P} \cos\delta \cos\alpha, 
\label{A2} \\
\tilde{B} &\equiv & \frac{1}{2}
 \left( 
 \left| B_{\sst \pi K}\right|^2 
 - \left| \bar{B}_{\sst \pi K}\right|^2 
 \right) = 
 - 2 \tilde{\cal T}\tilde{\cal P} 
 \sin\delta \sin\alpha, 
\label{B2} \\
\tilde{C} &\equiv & {\rm Im}\left( B_{\sst KK} 
 \bar{B}_{\sst KK}^* 
 \right)
 = r_u^2 \tilde{\cal T}^2 \sin2\gamma 
 - 2 r_u \tilde{\cal T }\tilde{\cal P}^\prime 
 \cos\delta \sin\gamma, 
\label{C2} \\ 
\tilde{D} &\equiv & \frac{1}{2}
 \left( 
 \left| B_{\sst KK}\right|^2 
 + \left| \bar{B}_{\sst KK}\right|^2 
 \right) = 
 r_u^2 \tilde{\cal T}^2 + \tilde{\cal P}^{\prime 2} 
 - 2 r_u \tilde{\cal T}
 \tilde{\cal P}^\prime
 \cos\delta \cos\gamma, 
\label{D2} \\ 
\tilde{E} &\equiv & \frac{1}{2}
 \left( 
 \left| B_{\sst KK}\right|^2 
 - \left| \bar{B}_{\sst KK}\right|^2 
 \right) = 
 2 r_u \tilde{\cal T}\tilde{\cal P}^\prime
 \sin\delta \sin\gamma, 
\label{E2} \\ 
\tilde{F} &\equiv & \left| B_{\sst KK}^s\right|^2 
 = {\tilde{{\cal P}}^{\prime 2}} 
\label{F2} 
\eea
Combined with the 6 quantities in Eqs.~(\ref{A1}-\ref{F1}), we have 12
equations in 9 unknowns. As shown below, this allows us to solve for the
CP angles, as in the DGR method, but greatly reduces the discrete
ambiguities.

The CP angles can be obtained as follows. First, one finds the ratios
$\tilde{\cal T}/{\cal T}$, $\tilde{\cal P}/{\cal P}$, and ${\tilde{\cal
P}}^\prime/{\cal P}^\prime$:
\beq
a \equiv \frac{\tilde{\cal T}} {\cal T} =
 \frac{\tilde{E}}{E}\sqrt{\frac{F}{\tilde{F}}} ~,~~~~
b \equiv \frac{\tilde{\cal P}}{\cal P} = 
 \frac{\tilde{B}E}{B \tilde{E}}\sqrt{\frac{\tilde{F}}{F}} ~,~~~~
c \equiv \frac{\tilde{\cal P}^\prime}{{\cal P}^\prime} = 
	 \sqrt{\frac{\tilde{F}}{F}} ~.
\label{abc}
\eeq
Using these, we can find the values of all the magnitudes of the
amplitudes. The amplitudes ${\cal T}$ and ${\cal P}$ are obtained from
\beq
{\cal T}^2 = \frac{( a c D - \tilde{D}) - c ( a - c ) F }
 { a ( c - a ) r_u^2 } ~,~~~~
{\cal P}^2 = \frac{a b A - \tilde{A}}{b ( a - b )} + 
 \frac{a}{b} \frac{(a c D - \tilde{D}) - 
 c ( c - a ) F }
 { a ( c - a ) r_u^2} ~,
\label{T2P2}
\eeq
and the remaining amplitudes can be found using Eq.~(\ref{abc}). Note that
all magnitudes are positive, by definition.

We now turn to the angles. Using our knowledge of the magnitudes of the
amplitudes, we have
\bea
\cos(\delta - \alpha ) &=& \frac{ {\cal T}^2 + {\cal P}^2 - A - B }
 { 2 {\cal T P }} ~, \nonumber \\
\cos(\delta + \alpha ) &=& \frac{ {\cal T}^2 + {\cal P}^2 - A + B }
 { 2 {\cal T P }} ~, \nonumber \\
\cos(\delta - \gamma ) &=& \frac{ r_u^2 {\cal T}^2 + F - D + E }
 { 2 r_u {\cal T} \sqrt{F}} ~, \nonumber \\
\cos(\delta + \gamma ) &=& \frac{ r_u^2 {\cal T}^2 + F - D - E }
 { 2 r_u {\cal T} \sqrt{F}} ~.
\label{cosines}
\eea
These equations can be solved to give the phases $\alpha$, $\gamma$ 
and $\delta$ up to a fourfold ambiguity. That is, if $\alpha_0$, $\gamma_0$
and $\delta_0$ are the true values of these phases, then the following
four sets of phases solve the above equations: $\{\alpha_0, \gamma_0,
\delta_0 \}$, $\{-\alpha_0, -\gamma_0, -\delta_0 \}$, $\{\alpha_0 - \pi,
\gamma_0 - \pi, \delta_0 - \pi\}$, and $\{\pi - \alpha_0, \pi - \gamma_0,
\pi - \delta_0 \}$. Note, however, that we still haven't used the $C$ and
$\tilde{C}$ measurements. Their knowledge eliminates two of the four sets,
leaving
\bea
&~& \{\alpha_0, \gamma_0, \delta_0\} ~, \nonumber \\
&~& \{\alpha_0 - \pi, \gamma_0 - \pi, \delta_0 - \pi\} ~.
\eea
These two solutions correspond to two different orientations of the
unitarity triangle, one pointing up, the other down. This final ambiguity
cannot be resolved by this method alone. However, within the SM it can be
removed by using other measurements such as $\epsilon $ in the kaon system
or the third CP angle $\beta$.

\section{More Comments}
For more details on this part of talk, 
please look the paper by Kim, London and Yoshikawa \cite{KLY}.

I also talked on several other topics: determinations of $|V_{td}/V_{ts}|$ from 
$B \to X_{s,d} l {\bar l}$, $(\sin \gamma /\sin \beta)$ from 
$B \to \rho(\pi) \nu {\bar \nu}$ and $|V_{ub}/V_{cb}|$ from invariant
hadronic mass distribution of $B \to X l {\bar \nu}$:
\begin{enumerate}

\item We propose\cite{KMS} a new method to extract ${|V_{td}| \over |V_{ts}|}$
from  the ratio of the decay distributions 
$ B \to X_d l {\bar l} / B \to X_s l {\bar l}$.
This ratio depends only on the KM ratio 
${|V_{td}| \over |V_{ts}|}$
within   $15\%$ theoretical uncertainties, 
if dilepton invariant mass-squared is away
from the peaks of the  possible resonance states,  
$J/\psi$, $\psi'$, and {\it etc.}
We also give a detailed analytical and numerical 
analysis on $ B \to X_q l {\bar l}$.

\item We propose\cite{AK} a new method for precise determination of $ \left |
\frac{V_{td}}{V_{ub}} \right | $ from the ratios of branching ratios $\frac{
{\cal B}(B \rightarrow \rho \nu \bar \nu )} { {\cal B}(B \rightarrow
\rho l \nu )}$ and $\frac{ {\cal B}(B \rightarrow \pi \nu \bar \nu )}
{ {\cal B}(B \rightarrow \pi l \nu )}$. 
As is well known, $ \left | \frac{V_{td}}{V_{ub}} \right | $ equals to
$ \left( \frac{\sin \gamma }{\sin \beta } \right)$ for the CKM version of
CP-violation within the Standard Model.

\item In order to determine the ratio of CKM matrix elements 
$|V_{ub}/V_{cb}|$ ~(and $|V_{ub}|$), we propose\cite{Kim} a new model-independent 
method based on the heavy quark effective theory.
In the forthcoming asymmetric $B$-experiments with microvertex 
detectors, BABAR and BELLE, the total separation of $b \rightarrow u$ 
semileptonic decays from the dominant 
$b \rightarrow c$ semileptonic decays would be experimentally viable.

\end{enumerate}

\section*{Acknowledgments}
I would like to thank the organizers of the first PPPP workshop and APCTP 
for their hard work to make the workshop successful.
The work is supported by the
Korean Reasearch Foundation made in the program year of 1997.

%\section*{References}

\end{document}